\documentclass[]{JHEP3}
\usepackage{amsmath,verbatim}

\title{Bogomol'nyi Bounds for Gravitational Cosmic Strings}
\author{Anne-Christine Davis and Senthooran Rajamanoharan\\
Department of Applied Mathematics and Theoretical Physics,
University of Cambridge \\
Wilberforce Road, Cambridge, CB3 0WA, United Kingdom\\
E-mail: \email{a.c.davis@damtp.cam.ac.uk},
\email{s.rajamanoharan@damtp.cam.ac.uk}}

\newcommand{\ud}{\mathrm{d}}
\newcommand{\cj}{\mathcal{J}}
\newcommand{\ssinf}{{\scriptscriptstyle\infty}}

\abstract{We present a new method for finding lower bounds on the
energy of topological cosmic string solutions in gravitational field theories.
This new method produces bounds that are valid over the entire space of
solutions, unlike the traditional approach, where the bounds obtained are
only valid for cylindrically symmetric solutions. This method is shown to be a
generalisation of the well-known Bogomol'nyi procedure for non-gravitational
theories and as such, it can be used to find gravitational Bogomol'nyi bounds
for models wherever the traditional Bogomol'nyi procedure can be applied in
the non-gravitational limit. Furthermore, this method yields Bogomol'nyi
equations that do not appear to rule out the existence of asymmetric
bound-saturating solutions.}

\preprint{DAMTP-2008-90}

\keywords{Solitons Monopoles and Instantons, Classical Theories of Gravity, Supergravity Models}

\begin{document}

\section{Introduction}

Topological defects are of considerable interest in many areas of 
theoretical and mathematical physics, and find application in topics
as diverse as superconductivity, nuclear physics and cosmology, as well as
having interesting mathematical properties in their own right. Whilst
most of these fields are concerned with the study of topological defects
in non-dynamical and often flat spacetimes,
within the field of cosmology it becomes important to examine the effect of
gravity on the properties and behaviour of these objects.

When studying topological defects, one is often interested in finding the
static configurations that minimise the total energy within each topologically
distinct class of boundary conditions. Such
configurations are the stable, classical ground states of the theory, and 
also form a convenient basis for numerical and analytical studies of low energy
defect dynamics.

In Minkowski space, or other non-dynamical, highly symmetrical spacetimes,
there is an established method, attributed to Bogomol'nyi, for finding
such minimum-energy field configurations in many models that admit solitonic
solutions (for a review, see~\cite{Manton:2004tk}). It
involves making use of a clever rearrangement of the energy-momentum tensor to
write the total energy as
\begin{equation}
E = \int\!\ud^{3}\mathbf{x}T^{0}_{\phantom{0}0} = |Q| + P~,
\end{equation}
where $Q$ is a topologically conserved charge, related to the asymptotic
boundary conditions, and $P$ is a manifestly non-negative spatial volume
integral. This leads to the following lower bound on the energy of
defects
\begin{equation}\label{eq:introbogbound}
E \geq |Q|~,
\end{equation}
which is called a Bogomol'nyi bound. The energy is minimised when $P$
is zero, and by finding the conditions under which $P$ vanishes, we obtain
a set of field equations, called Bogomol'nyi equations, that characterise the
minimum-energy field configurations.

However, for gravitational field theories, such energy bounds have been 
harder to come by, due to the difficulty in finding a suitable
expression for the total energy, and the more complicated form of
the energy-momentum tensor for a general metric.
One way forward is to reduce the number of degrees of freedom in the metric by
imposing certain exact symmetries on the spacetime \cite{Comtet:1987wi}.
When working within a well-chosen class of highly symmetric metrics, the
expressions for the total energy and the energy-momentum tensor become very
similar to their counterparts from the corresponding non-gravitational
theory, and therefore we can perform a similar rearrangement to minimise the
energy.

This approach is widely used
\cite{Vilenkin:1994,Dvali:2003zh,Burrage:2007bv}, and the energy
bounds and first-order equations that are derived in this manner are usually
called Bogomol'nyi bounds and equations. The beauty of this method is that it
is a natural extension of the familiar non-gravitational Bogomol'nyi method --
in fact, a gravitational bound of this sort follows wherever a similar bound
exists for the corresponding non-gravitational theory. This result, which
suggests that Bogomol'nyi bounds generally survive coupling
to gravity \cite{Achucarro:2004ry}, means that there is an implicit assumption
in much of the literature on gravitational topological defects that a
non-gravitational Bogomol'nyi bound is enough to establish the stability of
bound-saturating solutions even after gravity is taken into account
\cite{Davis:2005jf,Achucarro:2006ef}.

However, as the assumption of symmetry is made \emph{prior} to minimising
the energy, we cannot in fact preclude the possibility that the energy
bounds provided by this method may be saturated, or even violated, by
defects that do not possess the assumed symmetries. Intuitively speaking, we do
not expect such bound-violating solutions to exist -- we would be surprised to
find that non-gravitational Bogomol'nyi bounds do not survive coupling to
gravity. Nevertheless, without a more rigorous derivation of Bogomol'nyi bounds
for gravitational theories, the stability of the widely-studied defect
solutions that saturate these bounds is called into question.

Furthermore, it is often the case in non-gravitational theories that there
exist multi-defect solutions that saturate the Bogomol'nyi bounds, at least
for some region of the parameter space. Again,
it seems reasonable to consider whether such solutions survive the coupling to
gravity -- however, unless one can guess a sufficiently accurate ansatz for the
metric beforehand, the Bogomol'nyi technique of~\cite{Comtet:1987wi}
cannot help us answer this question.

For these reasons, it would be
worthwhile to pursue an alternative method for finding minimum-energy
solutions in gravitational field theories that does not depend on
making prior assumptions of symmetry. The pursuit of such a method would
involve tackling the problems mentioned above -- that of finding an
appropriate expression for the energy, and that of rearranging this expression
in the presence of a large number of degrees of freedom in the metric -- head
on.

The reader may have already noticed a striking resemblance between the
problem described here and the positive energy theorem in general relativity.
In fact, Witten's proof of the positive energy theorem
\cite{Witten:1981mf}, with its use of a spinorial expression for the total
energy, has already proved rather useful in establishing Bogomol'nyi bounds
for certain theories. Using techniques derived from
this proof, full Bogomol'nyi bounds have been constructed for certain
three-dimensional \cite{Edelstein:1995ba} and four-dimensional
\cite{Collinucci:2006sp} supergravity models with $D$-term symmetry
breaking.

In this paper, we extend these results and demonstrate how, using
techniques from the positive energy theorem, we may derive Bogomol'nyi
bounds for any gravitational field theory wherever a similar bound exists for
its non-gravitational counterpart. Due to the cosmological motivation for
this study, we only consider cosmic strings from now on. However, we
expect the methods presented here to be applicable, following appropriate
modifications, to defects of other dimensionalities, such as domain walls and
monopoles.

The rest of this paper is organised as follows. In Section~\ref{sec:css}
we describe the asymptotic structure of a spacetime containing a long cosmic
string, and consider how to express the total energy of such a spacetime.
We use this expression in Section~\ref{sec:gbogbounds} to
find a Bogomol'nyi bound for the gravitational version of the abelian-Higgs
model and
subsequently examine how this model forms the basis for finding
Bogomol'nyi bounds for many other field theories. Then in
Section~\ref{sec:comparison}, we compare our gravitational Bogomol'nyi
procedure to the traditional non-gravitational Bogomol'nyi procedure, and
show that the former is really a generalisation of the latter. In this way
we confirm that non-gravitational Bogomol'nyi bounds, and the single-vortex
solutions that saturate them, do survive coupling to gravity. We conclude in
Section~\ref{sec:conclusion}.

%%%%%%%%%%%%%%%%%%%%%%%%%%%%%%%%%%%%%%%%%%%%%%%%%%%%%%%%%%%%%%%%%%%%%%%%%%%%%
\section{Cosmic string spacetimes}\label{sec:css}

If we are to minimise the total energy of a spacetime without relying on
working within a class of highly symmetric metrics, then we must first
identify a suitable expression for the total energy. In general
relativity, the notion of total energy is closely tied up with the asymptotic
structure of the spacetime under consideration. This is because
the energy is a global quantity, dependent on the behaviour of the fields at
every point on some hypersurface that stretches to infinity. Therefore, without
being able to effectively compactify the spacetime, by specifying an
appropriate asymptotic structure, we cannot hope to calculate the total energy
of a system.

Where we have a compact source, the usual definitions of the energy (such
as the ADM energy) stem from the canonical notion of an asymptotically flat
spacetime~\cite{Wald:1984rg}. Such a spacetime can be compactified, with a
single point representing spatial infinity, and the spacetime becomes
asymptotically flat in every direction. However, it is clear that this
standard notion of asymptotic flatness is not appropriate for describing a
long cosmic string -- essentially because there is now an axial direction
(running parallel to the string) along which fields do not fall to zero and we
do not reach asymptotic flatness.

To resolve this problem, we must modify our notion of asymptotic flatness for a
cosmic string spacetime, by distinguishing between
radial infinity (where we do have asymptotic flatness) and
the asymptotic behaviour in the axial direction, on which we have to impose
suitable conditions in order to have a well-defined energy. We shall
accomplish this by compactifying one spatial dimension on a circle of
circumference $L_z$ and wrapping the cosmic string around this circle. In the
limit $L_z\to\infty$, edge effects should vanish, and the results we obtain
should reasonably represent the properties of an infinitely long string.

Having thus described the asymptotic structure of a cosmic string spacetime, we
note that, due to asymptotic flatness, there exists a neighbourhood
of radial infinity, the \emph{asymptotic region}, in which we can find
asymptotic cylindrical coordinates, $(t,r,\theta,z)$ for $r$ greater than some
constant $r_0$, in which the metric tends to the following limit as
$r\to\infty$:
\begin{equation}
\ud{s}^2 =
\ud{t}^2 - \ud{r}^2 - (1-\delta/2\pi)^{2}r^{2}\ud{\theta}^2-\ud{z}^2~,
\end{equation}
where $\delta$ is the conical deficit angle. These coordinates
shall turn out to be useful later on, when we examine the behaviour of fields
near radial infinity.
In order to fix the deficit angle $\delta$, we recall that the solutions
we are interested in should asymptotically tend to the static,
cylindrically symmetric bound-saturating solutions that have
already been found using the traditional Bogomol'nyi method of
\cite{Comtet:1987wi} (as these are the solutions that are relevant to
a discussion about the stability of static, cylindrically symmetric solutions).
For the static cylindrically symmetric solutions, one finds that
$\delta=2\pi|Q|$, where $Q$ is the topological charge of the string. Therefore,
we fix $\delta$ in a similar manner here.

We now turn to the question of how to define the total energy of a cosmic
string spacetime. In a \emph{canonical} asymptotically flat spacetime, the ADM
energy is defined with respect to some maximal spacelike hypersurface $S$
(i.e.~a spacelike hypersurface that extends to spatial infinity) in terms of a
surface integral over the asymptotic boundary of $S$, $\partial{S}$, at spatial
infinity. Given such an integral expression for the ADM energy, it seems
reasonable to speculate that the energy of a cosmic string spacetime can
be given by a similar expression, with the only difference being
that we replace spatial infinity by radial infinity, resulting in $\partial{S}$
having the topology of a torus, rather than a sphere.
If this is the case (as is confirmed in Appendix~\ref{sec:energy}), then in
order to find an expression for the energy of a cosmic string spacetime that
satisfies the requirements set out in the Introduction, we only need find an
appropriate expression for the ADM energy satisfying the same
conditions: we expect this expression to carry over to the cosmic string
spacetime following a simple change of the asymptotic surface of integration.

All that now remains is to identify a suitable expression for the
ADM energy -- one that, as described in the Introduction, is likely to admit
a Bogomol'nyi rearrangement for a fully general metric. In particular, it would
be ideal if the energy expression had a clear connection to the Minkowski
spacetime expression for the total energy, in terms of a volume integral
over the energy-momentum tensor.

Such an energy expression has been provided by Nester \cite{Nester:1982tr}
during his proof of the positive energy theorem:\footnote{Throughout this paper
we work in natural units, with $8\pi G=1$.}
\begin{equation}\label{eq:wnenergy}
p^{\mu}u^{\ssinf}_{\mu}=
  \frac{1}{2}\int_{\partial{S}}\!\!\!\ud{S}_{\mu\nu}E^{\mu\nu}~,
\end{equation}
where
\begin{equation}\label{eq:wnform}
E^{\mu\nu}=i\varepsilon^{\mu\nu\rho\sigma}
  \left(\bar{\eta}\gamma_{5}\gamma_{\rho}\nabla_{\sigma}\eta -
        \overline{\nabla_{\sigma}\eta}\gamma_{5}\gamma_{\rho}\eta \right)
\end{equation}
is the Witten-Nester 2-form.\footnote{The spinor covariant derivative is given
by $\nabla_{\mu}\eta = \partial_{\mu}\eta + 
\frac{1}{4}\omega_{\mu}^{\phantom{\mu}\underline{\nu\rho}}
\gamma_{\underline{\nu\rho}}\eta$, where $\gamma_{\mu\nu}=
\gamma_{[\mu}\gamma_{\nu]}$. Underlined indices are used
to represent frame (tetrad) components.}
The parameter $\eta$ is an arbitrary Dirac
spinor field that is asymptotically Killing ($\nabla_{\mu}\eta\to 0$), and
$u^{\mu}=\bar{\eta}\gamma^{\mu}\eta$ is hence an asymptotically constant,
timelike vector field. The asymptotic 4-vector $u^{\ssinf\mu}$ is the limit of
$u^{\mu}$ at spatial infinity, and represents the 4-velocity of the observer at
spatial infinity who is measuring the energy of the system.

Having converted this expression into a volume integral with the aid of the
divergence theorem, we apply the identity
\begin{equation}
\nabla_{[\mu}\nabla_{\nu]}\eta = -\frac{1}{8}
  R^{\underline{\rho\sigma}}_{\phantom{\underline{\rho\sigma}}\mu\nu}
   \gamma_{\underline{\rho\sigma}}\eta\ ,
\end{equation}
and Einstein's equation to find that
\begin{equation}\label{eq:wnenergyvol}
p^{\mu}u^{\ssinf}_{\mu} =
  \int_{S}\!\ud{S}_{\mu}\Big\{
    T^{\mu}_{\phantom{\mu}{\nu}}u^{\nu} - 
 2\overline{\nabla_{\nu}\eta}\gamma^{\nu\mu\rho}\nabla_{\rho}\eta
     \Big\}~.
\end{equation}

This integral is very similar to the Minkowski spacetime expression for the
energy. In fact, in a static spacetime, with $S$ normal to the timelike and
Killing $t$-direction, we can choose $\eta$ to be an exact Killing
spinor such that $u^{\mu}=(1,0,0,0)$ everywhere, and the above
expression reduces to
\begin{equation}\label{eq:wnenergyvolstatic}
p^{0} = \int_{S}\!\ud{V} T^{0}_{\phantom{0}{0}}~,
\end{equation}
which is exactly the Minkowski spacetime expression for the energy --
the very expression that is rearranged during the non-gravitational
Bogomol'nyi procedure. This is an encouraging sign that we may be able to
rearrange the Witten-Nester energy expression, in an analogous manner to the
non-gravitational rearrangement of~\eqref{eq:wnenergyvolstatic}, in order
to find Bogomol'nyi bounds in the presence of gravity.

%%%%%%%%%%%%%%%%%%%%%%%%%%%%%%%%%%%%%%%%%%%%%%%%%%%%%%%%%%%%%%%%%%%%%%%%%%%%%%%
\section{Bogomol'nyi bounds for gravitational field theories}
\label{sec:gbogbounds}

We shall now examine how the Witten-Nester energy expression may be used
to find Bogomol'nyi bounds for gravitational field theories in cosmic
string spacetimes.

Upon transferring the Witten-Nester energy expression to a cosmic string
spacetime, we immediately encounter a hitch: there are no globally
well-defined asymptotically Killing
spinors in a cosmic string spacetime with non-zero deficit angle.
This can be seen quite simply by noting that the $\theta$ component
of the Killing spinor equation is asymptotically
\begin{equation}
\partial_{\theta}\eta-\frac{1}{2} C' \gamma_{12}\eta = 0\ ,
\end{equation}
where $C' = 1-\delta/2\pi$. This has the general solution
\begin{equation}
\eta = \eta_{+} e^{\frac{i C' \theta}{2}} +
       \eta_{-} e^{\frac{-i C' \theta}{2}}\ ,
\end{equation}
where $\eta_{\pm}$ are coordinate-constant spinors satisfying the projection
conditions
\begin{equation}
(1\pm i\gamma_{12})\eta_{\pm}=0\ .
\end{equation}
Therefore, even if we set either $\eta_{+}$ or $\eta_{-}$ to zero, we can
only obtain a globally well-defined spinor $\eta$ if $C'=1$
(and hence $\delta=0$).

In order to circumvent this problem, let us now suppose that it is
possible to find some current, $\cj_{\mu}$, constructed from the matter fields,
such that
\begin{equation}\label{eq:asymcurrent}
\int_{S_{\infty}}\!\!\!\!\cj_{\mu}\ud{x}^{\mu} = 2\pi Q~,
\end{equation}
where $Q$ is the topological charge of the cosmic string and $S_\infty$ is
any closed curve at radial infinity that encircles the cosmic string once.
Using this current we can define a \emph{modified} covariant spinor derivative
$\hat{\nabla}_{\mu}$ by including an extra connection term as follows:
\begin{equation}\label{eq:modderiv}
\hat{\nabla}_{\mu}\eta = \nabla_{\mu}\eta + \frac{i}{2}\cj_{\mu}\eta~.
\end{equation}
We may now consider whether there exist any spinors that asymptotically
satisfy the \emph{modified} Killing spinor equation $\hat{\nabla}_{\mu}\eta=0$.
In fact, we can solve this equation asymptotically in a
similar manner to before, and now find the general asymptotic solution
\begin{equation}\label{eq:etaasymp}
\eta = \eta_{+} e^{\frac{i (C'-Q) \theta}{2}} +
       \eta_{-} e^{\frac{-i(C'+Q) \theta}{2}}\ .
\end{equation}
This solution \emph{is} globally well-defined, provided that
$\eta_{\mathrm{sign}(Q)}=0$.

We shall see that, for many models admitting solitonic string solutions,
a current $\cj_{\mu}$, satisfying condition \eqref{eq:asymcurrent} does exist.
For such models, we can therefore find asymptotically \emph{modified}-Killing
spinors, which satisfy the asymptotic projection condition
\begin{equation}\label{eq:projcond}
(1 + i\kappa\gamma_{12})\eta\to 0~,
\end{equation}
where $\kappa=\text{sign}(Q)$. We also note that this asymptotic
projection condition implies that, given an asymptotically modified-Killing
spinor $\eta$,  we can always find an asymptotic cylindrical
coordinate system in which the two asymptotically constant (and Killing)
vectors $u^{\mu}=\bar{\eta}\gamma_{\mu}\eta$ and
$v^{\mu}=\bar{\eta}\gamma_{5}\gamma_{\mu}\eta$ 
have the following limits as $r\to\infty$:
\begin{equation}
u^{\!\ssinf}_{\mu}\ud{x}^\mu = \ud{t}~,\qquad
v^{\!\ssinf}_{\mu}\ud{x}^\mu = \kappa\ud{z}~.
\end{equation}

Before proceeding, we ought to eliminate a potential source for confusion.
In supergravity theories, the notation $\hat{\nabla}_{\mu}$ is often used to
denote a particular choice of modified spinor derivative -- essentially one
where the current $\cj_{\mu}$ of our notation is identified with the gravitino
$U(1)$ connection $A^{B}_{\mu}$. Although, as we shall see later on, such a
choice allows us to define modified-Killing spinors for certain models, there
are other models in which the holonomy of $A^{B}_{\mu}$ no longer leads to the
cancellation in~\eqref{eq:etaasymp} that is required for the existence of
modified-Killing spinors. Furthermore, we would like our Bogomol'nyi procedure
to be just as applicable as its non-gravitational counterpart, which can be
applied to a model without regard to any supersymmetric extension the model
may or may not admit. For these reasons, we choose to define the
modified spinor derivative more generally, so that we can make a more
judicious choice of connection that allows for the existence of asymptotically
modified-Killing spinors.

Using this modified spinor derivative, we may define a modified
Witten-Nester 2-form $\hat{E}^{\mu\nu}$ in the following manner:
\begin{equation}\label{eq:mwnform}
\hat{E}^{\mu\nu}=i\varepsilon^{\mu\nu\rho\sigma}
  \left(\bar{\eta}\gamma_{5}\gamma_{\rho}\hat{\nabla}_{\sigma}\eta -
        \overline{\hat{\nabla}_{\sigma}\eta}\gamma_{5}\gamma_{\rho}\eta
    \right)~.
\end{equation}
Integrating $\hat{E}^{\mu\nu}$ over $\partial{S}$, we find
that the inclusion of the $\cj_\mu$ connection gives
\begin{equation}\label{eq:mwnint}
\frac{1}{2}\int_{\partial{S}}\!\!\!\ud{S}_{\mu\nu}\hat{E}^{\mu\nu}=
  \frac{1}{2}\int_{\partial{S}}\!\!\!\ud{S}_{\mu\nu}{E}^{\mu\nu}
  -2\pi\kappa Q L_z~,
\end{equation}
where $E^{\mu\nu}$ is the original Witten-Nester 2-form, defined as in
\eqref{eq:wnform}.

As argued in Appendix~\ref{sec:energy}, the integral of $E^{\mu\nu}$ gives
the energy of the cosmic string spacetime. Therefore,
dividing~\eqref{eq:mwnint} by $L_z$, we find that the energy per unit length,
$\mu$, satisfies
\begin{equation}\label{eq:mwnrear}
\mu = 2\pi|Q| +
  \frac{1}{2L_z}\int_{\partial{S}}\!\!\!\ud{S}_{\mu\nu}\hat{E}^{\mu\nu}~.
\end{equation}

Hence we can establish the Bogomol'nyi bound
\begin{equation}\label{eq:gbogbound}
\mu\geq2\pi|Q|~,
\end{equation}
if we can demonstrate that the integral of $\hat{E}^{\mu\nu}$ in
\eqref{eq:mwnrear} is non-negative. To this end, we repeat the manipulations
that took us from the surface integral \eqref{eq:wnenergy} to the volume
integral \eqref{eq:wnenergyvol}, and find that
\begin{equation}\label{eq:mwnvolint}
\frac{1}{2}\int_{\partial{S}}\!\!\!\ud{S}_{\mu\nu}\hat{E}^{\mu\nu} =
  \int_{S}\!\ud{S}_{\mu}\Big\{
    T^{\mu}_{\phantom{\mu}{\nu}}u^{\nu} - 
    \epsilon^{\mu\nu\rho\sigma}\partial_{\nu}\cj_{\rho}v_{\sigma} +
 2\overline{\hat{\nabla}_{\nu}\eta}\gamma^{\nu\mu\rho}\hat{\nabla}_{\rho}\eta
     \Big\}~.
\end{equation}
If we choose the $S$ to be normal to the $0$-direction, this becomes
\begin{equation}\label{eq:bogstart}
\frac{1}{2}\int_{\partial{S}}\!\!\!\ud{S}_{\mu\nu}\hat{E}^{\mu\nu} = 
  \int_{S}\!\ud{S}_{0}\Big\{
    T^{0}_{\phantom{0}{\nu}}u^{\nu} - 
    \epsilon^{0ijk}\partial_{i}\cj_{j}v_{k} - 
    2g^{ij}\hat{\nabla}_{i}\eta^{\dag}\hat{\nabla}_{j}\eta
    - 2\big(\gamma^{i}\hat{\nabla}_{i}\eta\big)^{\dag}
                          \big(\gamma^{j}\hat{\nabla}_{j}\eta\big)\Big\}~.
\end{equation}
The third term in this integral is manifestly positive-definite, whilst the
fourth term is negative-definite. However, the fourth term vanishes if the
spinor parameter $\eta$ satisfies
\begin{equation}\label{eq:mwittcond}
\gamma^{i}\hat{\nabla}_{i}\eta=0
\end{equation}
throughout $S$. In fact, as we shall demonstrate shortly, we can always choose
$\eta$ to satisfy this condition, as long as the inequality
\begin{equation}\label{eq:poscond}
T^{0}_{\phantom{0}{\nu}}u^{\nu} - 
    \epsilon^{0ijk}\partial_{i}\cj_{j}v_{k} \geq 0
\end{equation}
is satisfied throughout $S$. Therefore, the existence of a current $\cj_{\mu}$
that satisfies the inequality~\eqref{eq:poscond} is all that is required to
show that the integral~\eqref{eq:bogstart} is non-negative.

To summarise, we have found that the Bogomol'nyi
bound~\eqref{eq:gbogbound} can be established as long as we can find a current,
$\cj_{\mu}$, satisfying the asymptotic property~\eqref{eq:asymcurrent}, such
that the inequality~\eqref{eq:poscond} holds throughout $S$.

Let us now return to the condition~\eqref{eq:mwittcond}. This is a modified
version of the Witten-Nester condition, which was originally introduced by
Witten during his proof of the positive energy theorem \cite{Witten:1981mf}.
Adapting Witten's arguments, we shall now show that there always exists an
asymptotically modified-Killing spinor field $\eta$ that satisfies this
condition throughout $S$.

We begin by defining a spinor field $\eta_0$ that, in the asymptotic
region, takes the value
\begin{equation}
\eta_0 = \eta_{\kappa} e^{-\kappa\frac{i\theta}{2}}~,
\end{equation}
where $\eta_{\kappa}$ is a spinor, constant in cylindrical
coordinates, that satisfies the projection condition
\begin{equation}
(1+i\kappa\gamma_{12})\eta_{\kappa} = 0~.
\end{equation}
From~\eqref{eq:etaasymp}, we therefore see that $\eta_0$ is an asymptotically
modified-Killing spinor field. A more careful calculation, considering the
asymptotic fall-off rates of the metric and matter fields, shows that $\eta_0$
actually behaves as
\begin{equation}\label{eq:eta0asymp}
\gamma^{i}\hat{\nabla}_{i}\eta_{0} =
 \frac{\partial_{z}A(\theta,z)}{r}\eta_{0}
 + \mathcal{O}\left(\frac{1}{r^2}\right)~,
\end{equation}
for some function $A(\theta,z)$, defined on the torus at radial infinity.

Now, let us consider the inhomogeneous equation
\begin{equation}\label{eq:inhom}
\gamma^{i}\hat{\nabla}_{i}\eta_1 = -\gamma^{i}\hat{\nabla}_{i}\eta_0~,
\end{equation}
subject to the boundary condition that $\eta_1$ vanishes asymptotically.
It is straightforward to show that $\gamma^{i}\hat{\nabla}_{i}\eta=0$ has no
non-zero asymptotically vanishing solutions as long as the
inequality~\eqref{eq:poscond} is satisfied. Therefore, we can formally write
down the solution of~\eqref{eq:inhom} as
\begin{equation}
\eta_{1}(x) = \int_{S}\!\ud{y}G(x,y)\gamma^{i}\hat{\nabla}_{i}
 \left(-\gamma^{i}\hat{\nabla}_{i}\eta_{0}(y)\right)~,
\end{equation}
where $G(x,y)$ is the Green's function of the positive-definite, hermitian
second-order operator $-(i\gamma^{i}\hat{\nabla}_{i})^2$.

If this integral converges, it immediately follows that the spinor
$\eta=\eta_0 + \eta_1$ is both asymptotically modified-Killing, with limiting
value $\eta_0$, and also satisfies the modified Witten-Nester condition
throughout $S$.

To check the convergence of this integral, we perform a Fourier mode
expansion of the integrand along the circular $z$-direction.
From~\eqref{eq:eta0asymp}, it is clear that the zero-frequency component of the
source term $\gamma^{i}\hat{\nabla}_{i}\eta_0$ vanishes as $1/r^2$, whilst
all higher frequency components vanish as $1/r$. On the other hand,
the zero-frequency component of $G(x,y)$ grows logarithmically at
large distances, whilst other frequency components decay exponentially.
Putting these results together, we find that this integral is convergent, and
that $\eta_{1}$ asymptotically vanishes, at least as fast as
$(\mathrm{log}~r)/r$.

\subsection*{Bogomol'nyi bounds for the gravitational abelian-Higgs model}

In order to verify the inequality~\eqref{eq:poscond}, we need to identify a
suitable current $\cj_{\mu}$, which satisfies~\eqref{eq:asymcurrent}. Clearly,
the choice of a current $\cj_\mu$ that satisfies these conditions must be made
on a model-by-model basis, as this inequality depends on the form of the
energy-momentum tensor.

In fact, for non-gravitational models, a similar inequality,
\begin{equation}\label{eq:poscondcyl}
T^{0}_{\phantom{0}0} -
 \kappa\big(\partial_{r}J_{\theta}-\partial_{\theta}J_{r}\big) \geq 0~,
\end{equation}
is instrumental in establishing Bogomol'nyi bounds.
Therefore, for a given gravitational theory, it would be reasonable to
identify $\cj_{\mu}$ with the current $J_{\mu}$ that is
involved in establishing the Bogomol'nyi bound for the corresponding
non-gravitational theory. Having made this guess, we would then need to show
that this current both satisfies the condition~\eqref{eq:asymcurrent} and
enables us to establish the inequality~\eqref{eq:poscond}.

We shall begin by examining the gravitational abelian-Higgs model
\begin{equation}\label{eq:ahlag}
\mathcal{L}=
  \frac{1}{2}R+
  \hat{\partial}_{\mu}\phi^{*}\hat{\partial}^{\mu}\phi - 
  \frac{1}{4}F^{\mu\nu}F_{\mu\nu} \\ -
  \beta^{2}(\phi^{*}\phi-\xi)^2\ ,
\end{equation}
where $\phi$ is a $U(1)$-charged scalar field with covariant derivative
\begin{equation}
\hat{\partial}_{\mu}\phi=\partial_{\mu}\phi-igA_{\mu} \phi~,
\end{equation}
$g$ is the gauge coupling constant and $\xi$ is a
positive constant.

The non-gravitational limit of this theory, the abelian-Higgs model, is the
prototype for the traditional Bogomol'nyi procedure: it is the starting point
for the Bogomol'nyi rearrangement of the energy-momentum tensors of other
non-gravitational theories. Similarly,
we shall see that the Bogomol'nyi rearrangement of the gravitational
abelian-Higgs theory will enable us to obtain Bogomol'nyi bounds for a 
variety of gravitational theories.

In the (non-gravitational) abelian-Higgs model, the current
\begin{equation}\label{eq:ahcurrent}
J_{\mu}=\frac{i}{2} 
 \big[\phi\big(\hat{\partial}_{\mu}\phi\big)^{*}-
     \phi^{*}\big(\hat{\partial}_{\mu}\phi\big)\big] + g\xi A_{\mu}~,
\end{equation}
satisfies the inequality~\eqref{eq:poscondcyl} and therefore
establishes a Bogomol'nyi bound. $J_\mu$ also provides us with the topological
charge $Q=n\xi$, due to the boundary conditions satisfied by finite-$\mu$
field configurations. Following our earlier discussion, we therefore
make the identification $\cj_{\mu} = J_{\mu}$. It is straightforward to check
that the same current produces the conserved topological charge $Q=n\xi$
in the gravitational abelian-Higgs theory. Therefore, we now turn to
proving the inequality~\eqref{eq:poscond} for this choice of current.

The key to verifying this
inequality is to notice that, when the parameters $\beta$ and $g$ are in the
Bogomol'nyi limit ($\beta^{2}=g^{2}/2$), the gravitational abelian-Higgs
model is the bosonic limit of an $N=1$ supergravity theory with
$D$-term symmetry breaking -- a model with a single
charged chiral superfield, simple K\"ahler
potential, a non-zero Fayet-Iliopoulos constant $\xi$ and a superpotential
that is identically zero. Furthermore, our choice of $\cj_{\mu}$ coincides
with the gravitino $U(1)$ connection $A^{B}_{\mu}$.

A gravitational Bogomol'nyi bound has
already been established for this theory in~\cite{Collinucci:2006sp},
where it was noticed that the total energy for this system could be written as
the sum of squares of the fermionic supersymmetry transformations.
In terms of the formalism described here, this is equivalent to showing
that the left-hand side of the inequality \eqref{eq:poscond} can be written
as a sum of squares of certain spinorial quantities, defined as follows:
\begin{align}\label{eq:chidef}
\delta\chi &=
   -\frac{i}{2}\gamma^{\mu}\big(\hat{\partial}_{\mu}\phi\big)\eta~,\\
\delta\lambda &=
   -\frac{i}{4}\gamma^{\mu\nu}F_{\mu\nu}\eta
   - \frac{g}{2}(\phi^{*}\phi-\xi)\eta~.
\label{eq:lambdadef}
\end{align}
As the notation suggests, these quantities are clearly related to the
higgsino and gaugino supersymmetry transformations respectively. In fact, each
is a linear combination of supersymmetry transformations given by the two Weyl
components that are encoded in the Dirac spinor $\eta$. Furthermore, with
$\cj_{\mu}$ defined as in \eqref{eq:ahcurrent}, $\hat{\nabla}_{\mu}\eta$
is a linear combination of gravitino supersymmetry transformations in
the same manner.

With a little effort, one can show that
\begin{equation}\label{eq:gahid1}
4\overline{\delta\chi}\gamma^{\mu}
 \delta\chi = \Big[
   \hat{\partial}^{\mu}\phi^{*}\hat{\partial}_{\nu}\phi+
   \hat{\partial}_{\nu}\phi^{*}\hat{\partial}^{\mu}\phi-
   \delta^{\mu}_{\phantom{\mu}\nu}
     \hat{\partial}^{\rho}\phi^{*}\hat{\partial}_{\rho}\phi
   \Big]u^{\nu}
  -\epsilon^{\mu\nu\rho\sigma}\partial_{\nu}\cj_{\rho}v_{\sigma}
   -\frac{g}{2}(\phi^{*}\phi-\xi)\epsilon^{\mu\nu\rho\sigma}
                                              F_{\nu\rho}v_{\sigma}~,
\end{equation}
and
\begin{equation}\label{eq:gahid2}
2\overline{\delta\lambda}\gamma^{\mu}\delta\lambda
 =\Bigg[ F^{\mu\rho}F_{\rho\nu} - \delta^{\mu}_{\phantom{\mu}\nu}
   \bigg(\!-\!\frac{1}{4}F^{\rho\sigma}F_{\rho\sigma}-\frac{g^2}{2}
    (\phi^{*}\phi-\xi)\bigg)\Bigg]u^{\nu}
  +\frac{g}{2}(\phi^{*}\phi-\xi)\epsilon^{\mu\nu\rho\sigma}
                                              F_{\nu\rho}v_{\sigma}~.
\end{equation}
Hence we can rewrite the left-hand side of \eqref{eq:poscond} as a sum
of squares:
\begin{equation}
T^{0}_{\phantom{0}{\nu}}u^{\nu} - 
    \epsilon^{0ijk}\partial_{i}\cj_{j}v_{k} = 
4\delta\chi^{\dag}\delta\chi + 
2\delta\lambda^{\dag}\delta\lambda
+\bigg(\beta^2 - \frac{g^2}{2}\bigg)(\phi^{*}\phi-\xi)^2~.
\end{equation}

Therefore, provided that $\beta^{2}\geq g^{2}/2$, we obtain the Bogomol'nyi
bound
\begin{equation}\label{eq:gahbogbound}
\mu \geq 2\pi|n|\xi~.
\end{equation}
In the Bogomol'nyi limit $\beta^{2} = g^{2}/2$, this bound is saturated
when each positive-definite term in \eqref{eq:bogstart} vanishes throughout
$S$. This yields the following Bogomol'nyi equations:
\begin{equation}\label{eq:gahbeqs}
\delta\chi = \delta\lambda =
\hat{\nabla}_{i}\eta = 0~.
\end{equation}

Notice that the Bogomol'nyi equation for $\eta$ implies that $u^{\mu}$ and
$v^{\mu}$ are constant (and Killing) throughout $S$. The
existence of these two Killing vectors implies that minimum-energy cosmic
string solutions are static and translationally invariant along
the $z$-axis (cf.~the traditional gravitational Bogomol'nyi method,
following~\cite{Comtet:1987wi}, where these symmetries were assumed, rather
than derived). Furthermore, regarding this theory as a $D$-term supergravity
model, solutions of the Bogomol'nyi equations~\eqref{eq:gahbeqs} partially
preserve supersymmetry -- i.e.~they are BPS solutions.

\subsection*{Bogomol'nyi bounds for other gravitational field theories}

Having obtained a Bogomol'nyi bound and Bogomol'nyi equations for the
gravitational abelian-Higgs theory, it is now possible to construct Bogomol'nyi
bounds and equations for other gravitational field theories.

This is achieved
by noticing that any symmetry-breaking term of the form
\begin{equation}\label{eq:sbtermgen}
\beta^2\big[\mathcal{M} - \xi\big]^2~,
\end{equation}
where $\mathcal{M}$ is a real-valued quadratic form with respect to the scalar
fields $\phi_i$ and their complex conjugates $\phi_i^*$, can be brought to the
form
\begin{equation}
\tilde{\beta}^2\bigg(\sum_{i}q_{i}|\psi_{i}|^{2}-\tilde{\xi}\bigg)^2~,
\end{equation}
where $\psi_{i}$ are a suitably chosen (charge-preserving) unitary
transformation of the fields $\phi_{i}$, and $q_i$ are the charges of
the fields $\psi_i$. Furthermore, such a transformation will leave the
kinetic terms in the energy-momentum tensor unchanged:
\begin{equation}
\sum_{i}|\partial\phi_i|^2=\sum_{i}|\partial\psi_i|^{2}~.
\end{equation}
Therefore, this
field transformation effectively turns any theory that contains a symmetry
breaking potential of the form \eqref{eq:sbtermgen} into an abelian-Higgs
theory (perhaps with some extra terms in the scalar potential). Hence we can
minimise the energy by applying the abelian-Higgs Bogomol'nyi
rearrangement to the energy-momentum tensor, written in terms
of the new fields $\psi_i$.

As a concrete example, let us consider a popular model from $N=1$ supergravity
-- an $F$-term symmetry-breaking model containing three
chiral superfields $\Phi_0$ and $\Phi_\pm$, with charges $0$ and $\pm 1$, and
the superpotential
\begin{equation}\label{eq:ftsup}
W = \beta \Phi_0 \bigg(\Phi_{+}\Phi_{-} - \frac{\xi_F}{2}\bigg)~,
\end{equation}
where $\xi_F$ is a positive constant.
This superpotential gives rise to the following symmetry-breaking scalar
potential:
\begin{equation}
V=\alpha^{2}(\text{Re }F)^2+V_{\text{rest}}~,
\end{equation}
where
\begin{align}
\alpha^2 &= \beta^{2}\big[(1-|\phi_0|^{2})^{2}+|\phi_0|^2\big]
 e^{\sum_{i}\!|\phi_{i}\!|^2}~,\\
F &= \phi_{+}\phi_{-}-\xi_F/2~,
\end{align}
and
\begin{equation}
V_{\text{rest}}=\beta^{2}|\phi_0|^{2}e^{\sum_{i}\!|\phi_{i}\!|^2}
            \Big[|\phi_{-}+\phi_{+}^{*}F|^2+|\phi_{+}+\phi_{-}^{*}F|^2\Big]
 +\alpha^{2}(\text{Im }F)^2
 +\frac{g^2}{2f}\Big(|\phi_+|^2-|\phi_-|^2\Big)^2~.
\end{equation}
The function $f$ is the gauge kinetic function, which appears in the
bosonic Lagrangian as follows:
\begin{equation}\label{eq:ftlag}
\mathcal{L}=\frac{1}{2}R
       +\sum_{i}\hat{\partial}_{\mu}{\phi_i}^{*}\hat{\partial}^{\mu}\phi_{i}
       -\frac{f}{4}F^{\mu\nu}F_{\mu\nu}-V~.
\end{equation}
We will leave $f$ unspecified for now, so that we can examine how our choice
of $f$ affects the existence, and attainability, of a Bogomol'nyi bound for
this model.

We shall now consider topological cosmic strings in the bosonic limit of this
model. The scalar potential $V$ is manifestly non-negative, and takes the
minimum value of zero when $\phi_0 = 0$, $|\phi_+|=\sqrt{\xi}$ and
$\phi_{-}=\phi_{+}^{*}$. Therefore the vacuum manifold has a $U(1)$ topology,
and this model admits topological cosmic string configurations.

If we now rotate to the fields $\psi_0$ and $\psi_{\pm}$, where
\begin{equation}\label{eq:ftfieldredef}
\psi_{0}= \phi_{0}\quad\text{and}\quad
\psi_{\pm} = \frac{1}{\sqrt2}\big(\phi_{\pm} \pm \phi_{\mp}^{*}\big)~,
\end{equation}
then we find that $V$ becomes
\begin{equation}
V=\frac{\alpha^2}{4}\Big(|\psi_+|^2-|\psi_-|^2-\xi_F\Big)^2 + V_{\text{rest}}~,
\end{equation}
whilst the kinetic terms in the energy-momentum tensor are unchanged.
Therefore, written in terms of the new fields $\psi_0$ and $\psi_\pm$, this
$F$-term model has an energy-momentum tensor which is the sum of the 
abelian-Higgs energy-momentum tensor and the scalar potential
$V_{\text{rest}}$.

Hence, by generalising $\delta\chi$ and $\delta\lambda$ as follows
\begin{align}\label{eq:chidefg}
\delta\chi_i &=
   -\frac{i}{2}\gamma^{\mu}\big(\hat{\partial}_{\mu}\psi_{i}\big)\eta~,\\
\delta\lambda &=
   -\frac{i}{4}\gamma^{\mu\nu}F_{\mu\nu}\eta
   - \frac{g}{2}\bigg(\sum_{i}q_{i}|\psi_{i}|^2-\xi_F\bigg)\eta~,
\label{eq:lambdadefg}
\end{align}
and using the current
\begin{equation}\label{eq:ahcurrentg}
\cj_{\mu}=\frac{i}{2} \sum_{i}q_{i}
 \big[\psi_{i}\big(\hat{\partial}_{\mu}\psi_{i}\big)^{*}-
     {\psi_i}^{*}\big(\hat{\partial}_{\mu}\psi_{i}\big)\big] + g\xi_F A_{\mu}~,
\end{equation}
which can easily be shown to satisfy the condition~\eqref{eq:asymcurrent},
thereby enabling the existence of asymptotically modified-Killing spinors, we
find that
\begin{equation}
T^{0}_{\phantom{0}{\nu}}u^{\nu} - 
    \epsilon^{0ijk}\partial_{i}\cj_{j}v_{k} =
\sum_{i}4\delta{\chi_i}^{\dag}\delta\chi_i + 
2\delta\lambda^{\dag}\delta\lambda + 
\frac{1}{4f}(\alpha^{2}f - 2g^{2})
 \bigg(\sum_{i}q_{i}|\psi_i|^2-\xi_F\bigg)^2 +
V_{\text{rest}}~.
\end{equation}
Substituting this into \eqref{eq:bogstart}, we obtain the Bogomol'nyi
bound
\begin{equation}
\mu \geq 2\pi|n|\xi_{F}~,
\end{equation}
provided that $f$ satisfies the inequality $\alpha^{2}f\geq 2g^2$.

For the usual choice $f=1$, this inequality is satisfied as long as
$\beta^2=2g^2$, since $\alpha^2 \geq 1$ everywhere. However, the inequality
cannot be saturated everywhere, and hence this Bogomol'nyi bound cannot
be saturated either.

For this Bogomol'nyi bound to be attainable, we need to choose $f$ according
to the formula $\alpha^{2}f=2g^2$. This corresponds to the Bogomol'nyi
limit, or critical coupling, in the abelian-Higgs model, where we had to relate
the gauge coupling constant to the mass of the scalar field in order to obtain
an attainable bound. For this choice of $f$, this Bogomol'nyi bound is
saturated by field configurations that satisfy the following Bogomol'nyi
equations:
\begin{equation}\label{eq:ftbeqs}
\delta\chi_i = \delta\lambda =
\hat{\nabla}_{i}\eta = V_{\text{rest}} = 0~.
\end{equation}
We note that, just as the non-gravitational version of this model has
embedded Nielsen-Olesen strings as minimum-energy solutions, the above
Bogomol'nyi equations are solved by embedded minimum-energy solutions
of the gravitational abelian-Higgs model.

Let us now consider the relationship between these results and supersymmetry.
The Lagrangian we have just considered is derived from the bosonic limit of
an $F$-term $N=1$ supergravity model. For this model, with its non-vanishing
superpotential, it is easily seen that the BPS equations cannot be
satisfied~\cite{Brax:2006yb} -- i.e.~there are no non-trivial minimum-energy
configurations that preserve any degree of
supersymmetry.\footnote{In this paper, we take Bogomol'nyi equations to be the
equations that characterise minimum-energy solutions, whilst BPS equations are
the equations that characterise partially supersymmetric field configurations.
Although these two properties often come hand in hand, this is not the case
here, and we must therefore distinguish between the equations that
characterise these two properties.}
Therefore, it may initially seem rather surprising that we have been able to
obtain an energetic Bogomol'nyi bound for this Lagrangian. However, on closer
inspection, this result turns out to be a consistent generalisation of
analogous results for $F$-term strings in both (global)
supersymmetry and supergravity in the cylindrically symmetric limit -- that,
although the BPS equations cannot be satisfied, one can still establish an
energetic Bogomol'nyi bound~\cite{Dvali:2003zh}.

The relationship between our energetic Bogomol'nyi bound and the
non-existence of supersymmetric solutions manifests itself in a number of ways.
Firstly, it is clear that the Bogomol'nyi equations~\eqref{eq:ftbeqs} are not
equivalent to the $F$-term BPS equations. Secondly, the Bogomol'nyi bound is
only attainable when $\alpha^2 f=2g^2$ -- a choice which would result in $f$
not being holomorphic, and therefore not a valid choice if~\eqref{eq:ftlag} is
to be the bosonic part of a supergravity Lagrangian. Furthermore, and most
importantly, the current $\cj_{\mu}$ is not the gravitino $U(1)$ connection
$A^{B}_{\mu}$ -- a key result that enabled us to find an appropriate spinor
parameter $\eta$ for the Witten-Nester energy when Killing spinors, in the
usual supergravity sense, cannot exist for $\delta>0$.

%%%%%%%%%%%%%%%%%%%%%%%%%%%%%%%%%%%%%%%%%%%%%%%%%%%%%%%%%%%%%%%%%%%%%%%%%%%%%%%
\section{Comparison to Bogomol'nyi bounds for non-gravitational
theories}\label{sec:comparison}

There is a strong analogy between the gravitational Bogomol'nyi method
we have presented here and the traditional Bogomol'nyi procedure for
vortices in non-gravitational models. In fact, it is more appropriate to say
that our new method is really a generalisation of the non-gravitational
Bogomol'nyi method.

This analogy begins with the Witten-Nester energy expression which, as
mentioned earlier, is the gravitational energy expression that is closest,
for our purposes, to the definition of the total energy in Minkowski spacetime.
Using the Witten-Nester energy, we were able to come up with an inequality
\eqref{eq:poscond}, that must be satisfied in order to establish a
gravitational Bogomol'nyi bound. As we saw there, this inequality is
a generalised version of the inequality \eqref{eq:poscondcyl} that is
obtained by the non-gravitational Bogomol'nyi procedure.

Furthermore, the identities \eqref{eq:gahid1} and \eqref{eq:gahid2} are
generalisations of the Minkowski spacetime identities
\begin{equation}\label{eq:musefulid}
 \bigg|\hat{\partial}_r\phi\pm\frac{i}{r}\hat{\partial}_{\theta}\phi\bigg|^2=
 \big|\hat{\partial}_r\phi\big|^2+
 \frac{1}{r^2}\big|\hat{\partial}_{\theta}\phi\big|^2+
 \mp\frac{1}{r}(\partial_{r}\cj_{\theta}-\partial_{\theta}\cj_{r}) 
 \mp\frac{g}{r}F_{r\theta}(\phi^{*}\phi-\xi) ~,
\end{equation}
and
\begin{equation}
\frac{1}{2}\bigg(\frac{F_{r\theta}}{r}\pm g(\phi^{*}\phi-\xi)\bigg)^2 = 
 \frac{{F_{r\theta}}^2}{2r^2}+\frac{g^2}{2}(\phi^{*}\phi-\xi)^2 \pm
 \frac{g}{r}F_{r\theta}(\phi^{*}\phi-\xi)~,
\end{equation}
that are used in the non-gravitational Bogomol'nyi method to write the
left-hand side of \eqref{eq:poscondcyl} as a sum of squares.

Finally, the field transformation technique discussed above -- which relates
many gravitational field theories to the gravitational abelian-Higgs model, and
therefore allows us to construct Bogomol'nyi bounds for these theories -- also
relates the Bogomol'nyi bounds and equations of the non-gravitational
versions of these theories to the non-gravitational abelian-Higgs model.

This correspondence between the gravitational Bogomol'nyi method
presented here and the traditional Bogomol'nyi method for non-gravitational
theories demonstrates that non-gravitational Bogomol'nyi bounds do
survive coupling to gravity. Wherever we can construct
a Bogomol'nyi bound for a non-gravitational theory, we can use the
techniques described above to generalise this Bogomol'nyi rearrangement, and
therefore provide a Bogomol'nyi bound for the gravitational version of the
same theory.

In this manner, we can establish Bogomol'nyi bounds for many other
gravitational theories that are currently of cosmological interest -- such as
$P$-term models~\cite{Burrage:2007bv},
semi-local models~\cite{Achucarro:1999it}, and $D$-term models with non-zero
superpotentials -- where Bogomol'nyi rearrangements are already known
in the non-gravitational limit.

%%%%%%%%%%%%%%%%%%%%%%%%%%%%%%%%%%%%%%%%%%%%%%%%%%%%%%%%%%%%%%%%%%%%%%%%%%%%%%%
\section{Conclusion}\label{sec:conclusion}

We have presented a general method for establishing Bogomol'nyi bounds and
finding minimum-energy cosmic string solutions that can be applied to a wide
range of gravitational field theories that contain symmetry-breaking scalar
potentials. Unlike the traditional method for establishing energy bounds for
gravitational theories \cite{Comtet:1987wi}, this new method does not involve
making any prior assumptions about the symmetries of minimum-energy
solutions.

Our work generalises the results of~\cite{Edelstein:1995ba} and
\cite{Collinucci:2006sp}, regarding certain $D$-term supergravity models.
Although the algebraic manipulations that enabled us to derive these
bounds were borrowed from $D$-term supergravity, they were actually found to
be applicable to a wide variety of (possibly non-supersymmetric) theories.

In fact we have seen, in Section~\ref{sec:comparison}, that these
manipulations, although taken from a supersymmetric context, are really the
covariant generalisations of the key identities that were used to derive the
non-gravitational Bogomol'nyi bound for the abelian-Higgs model. In this sense,
our procedure is really a covariant generalisation of the traditional
Bogomol'nyi procedure. Therefore, we can confirm that all results
that have been proven so far by making assumptions about the symmetry of the
metric and using the traditional Bogomol'nyi technique, still hold when one
allows for asymmetric perturbations.

We applied our technique to the particular example of the bosonic Lagrangian
for  $F$-term strings in $N=1$ supergravity. Here, we found an energetic
Bogomol'nyi bound and corresponding Bogomol'nyi equations -- however, this
bound is unnattainable for any holomorphic choice of gauge kinetic function.
These results, which have been derived for cylindrically symmetric strings
in~\cite{Dvali:2003zh}, have therefore been generalised by this technique to
cover cylindrically asymmetric field configurations.

The Bogomol'nyi equations obtained using this new
technique confirm that minimum-energy solutions are static and
straight -- symmetries that were previously assumed rather than proved.
Furthermore these equations allow for the same
cylindrically symmetric single-vortex solutions as the Bogomol'nyi equations of
the traditional method, thereby confirming the stability of these
solutions against decay to asymmetric field configurations of lower energy.

However, the new Bogomol'nyi equations do not necessarily imply that
minimum-energy solutions must be cylindrically symmetric. Therefore, it
would be interesting to look for multi-string configurations that
saturate the Bogomol'nyi bound, in analogy with the
static multi-vortex solutions that exist in non-gravitational field theories.
To find such solutions, one would presumably have to repeat Taubes's analysis
of the abelian-Higgs Bogomol'nyi equations~\cite{Taubes:1979tm} for the
gravitational Bogomol'nyi equations~\eqref{eq:gahbeqs}.

\section*{Acknowledgements}

We would like to thank P.~Brax and N.~S.~Manton for useful
discussions. This work is supported by the UK Science and Technology
Facilities Council (STFC).

%%%%%%%%%%%%%%%%%%%%%%%%%%%%%%%%%%%%%%%%%%%%%%%%%%%%%%%%%%%%%%%%%%%%%%%%%%%%%%
\appendix

\section{The energy of cosmic string spacetimes}
\label{sec:energy}

In Section~\ref{sec:css} we claimed that any surface integral expression for
the energy of a canonical asymptotically flat spacetime can also be used
to calculate the energy of a cosmic string spacetime, provided that the surface
of integration is changed from the sphere at spatial infinity to the
torus at radial infinity. Then, in Section~\ref{sec:gbogbounds}, we employed
this result to interpret the surface integral of $E^{\mu\nu}$.

There are many ways to check this assertion. One would be to see
whether an energy expression obtained in this manner is equivalent to the
gravitational Hamiltonian obtained by following the background-subtraction
procedure described by Hawking and Horowitz \cite{Hawking:1995fd}. However,
here we shall adopt a more pedagogical approach by employing the principles
that were used to define the ADM energy of canonical asymptotically flat
spacetimes to define an equivalent energy for cosmic string spacetimes.

We begin by considering linearised gravity on a static, cylindrically symmetric
background. More specifically, let us consider a spacetime that admits global
cylindrical polar coordinates, in which the metric may be written
\begin{equation}
g_{\mu\nu}=\tilde{g}_{\mu\nu}+\epsilon h_{\mu\nu}\ ,
\end{equation}
where
\begin{equation}\label{eq:csmetric}
\tilde{g}_{\mu\nu}\ud{x}^{\mu}\ud{x}^{\nu} = 
 \ud{t}^{2}-\ud{r}^{2}-C(r)^{2}\ud{\theta}^{2}
 -\ud{z}^{2}\
\end{equation}
is a static, cylindrically symmetric background metric, and $\epsilon$ is a
small parameter, with respect to which we shall linearise all quantities. We
shall also assume that $C'(0)=0$ and $C(r)\to (1-\frac{\delta}{2\pi})r$ as
$r\to \infty$,
so that the background metric is completely regular, with deficit angle
$\delta$.

In the linearised theory, the dynamics of the matter fields take place with
respect to the fixed background (to leading order), and therefore the energy
may be defined in a special-relativistic sense, in terms of the linearised
energy-momentum tensor:
\begin{equation}\label{eq:srenergy}
p^{\mu}u_{\mu}=\int_{S}\!\ud^{3}x \sqrt{-\tilde{g}}
  {T^{(L)0}}_{\phantom{0}\nu}u^{\nu}\ ,
\end{equation}
where $S$ is a background-static, spacelike hypersurface whose normal vector
points in the (timelike and background-Killing) $t$-direction, and $u^{\mu}$
is a constant, timelike vector that represents the $4$-velocity of the
observer at spacelike infinity who is measuring the energy of the system.
Due to Einstein's equation, we can replace
$T^{(L)\mu}_{\phantom{(L)\mu}\nu}$ with $G^{(L)\mu}_{\phantom{(L)\mu}\nu}$,
the linearised Einstein tensor, in the above expression.

It is possible to express the linearised Riemann tensor
$R^{(L)\mu\nu}_{\phantom{(L)\mu\nu}\rho\sigma}$ in terms of the background
Riemann tensor $\tilde{R}^{\mu\nu}_{\phantom{\mu\nu}\rho\sigma}$, the
background metric connection $\tilde{\Gamma}^{\mu\nu}_{\phantom{\mu\nu}\rho}$,
and the linearised perturbation of the spin connection
$\Delta\omega_{\rho}^{\phantom{\rho}\mu\nu}$, in the following manner:
\begin{equation}
R^{(L)\mu\nu}_{\phantom{(L)\mu\nu}\rho\sigma} =
  \tilde{R}^{\mu\nu}_{\phantom{\mu\nu}\rho\sigma} + 2\left(
    \Delta\omega^{\phantom{[\rho}\mu\nu}_{[\rho\phantom{\mu\nu},\sigma]} -
    2\Delta\omega^{\phantom{[\rho}\tau[\mu}_{[\rho}
     \tilde{\Gamma}^{\nu]}_{\phantom{\nu]}\sigma]\tau}
  \right)\ .
\end{equation}
With the aid of this expression, along with the identity
\begin{equation}
G^{\mu}_{\phantom{\mu}\nu}=\frac{1}{4}\epsilon^{\rho\mu\gamma\delta}
  \epsilon_{\rho\nu\alpha\beta}
  R^{\alpha\beta}_{\phantom{\alpha\beta}\gamma\delta}\ ,
\end{equation}
we can rewrite \eqref{eq:srenergy} as follows:
\begin{equation}\label{eq:csenergy}
p^{\mu}u_{\mu}=2\pi\delta L_z+\frac{1}{4}
 \int_{\partial S}\!\!\!\ud S_{\mu\nu}
  \varepsilon^{\mu\nu\rho\sigma}\varepsilon_{\delta\alpha\beta\sigma}
  \Delta\omega_{\rho}^{\phantom{\rho}\alpha\beta}u^{\delta}~.
\end{equation}
We note that this expression is formally equivalent to Nester's
expression of the ADM energy in~\cite{Nester:1982tr}.

Now that we have an expression for the energy which only depends on the
\emph{asymptotic} behaviour of the system, we may allow $u^{\mu}$ to
be any \emph{asymptotically} constant, timelike vector field. Similarly, we
may now allow $S$ to be any maximal spacelike hypersurface, with a timelike,
\emph{asymptotically} Killing normal vector.
In this manner, we obtain an expression for the energy which
only depends on the asymptotic form of the metric.

Now we invoke the canonical argument that, on physical grounds, we
require the mass of a system to be determined purely by the long-distance
behaviour of the metric, and hence be independent of the fields in the
interior of the system. This implies that the expression \eqref{eq:csenergy}
represents the total energy of \emph{any} spacetime that is asymptotically
cylindrically symmetric, irrespective of its behaviour in the interior.
In other words,~\eqref{eq:csenergy} represents the energy of any cosmic
string spacetime.

It is straightforward to check that the surface integral of $E^{\mu\nu}$ equals
this energy expression, by decomposing the spin connection in the asymptotic
region as
\begin{equation}
\omega_{\mu}^{\phantom{\mu}\underline{\alpha\beta}} = 
 \tilde{\omega}_{\mu}^{\phantom{\mu}\underline{\alpha\beta}} +
 \Delta\omega_{\mu}^{\phantom{\mu}\underline{\alpha\beta}}\ ,
\end{equation}
where $\tilde{\omega}_{\mu}^{\phantom{\mu}\underline{\alpha\beta}}$ is the
spin connection for a cylindrically symmetric metric of identical conical
deficit angle.

\bibliography{main}

\providecommand{\href}[2]{#2}\begingroup\raggedright\begin{thebibliography}{10}

\bibitem{Manton:2004tk}
N.~S. Manton and P.~Sutcliffe, {\em Topological Solitons}.
\newblock Cambridge Monographs on Mathematical Physics. CUP, 2004.

\bibitem{Comtet:1987wi}
A.~Comtet and G.~W. Gibbons, {\it Bogomol'nyi bounds for cosmic strings},  {\em
  Nucl. Phys.} {\bf B299} (1988) 719.

\bibitem{Vilenkin:1994}
A.~Vilenkin and E.~P.~S. Shellard, {\em Cosmic Strings and other Topological
  Defects}.
\newblock Cambridge Monographs on Mathematical Physics. CUP, 1994.

\bibitem{Dvali:2003zh}
G.~Dvali, R.~Kallosh, and A.~Van~Proeyen, {\it {D}-term strings},  {\em JHEP}
  {\bf 01} (2004) 035, [\href{http://xxx.lanl.gov/abs/hep-th/0312005}{{\tt
  hep-th/0312005}}].

\bibitem{Burrage:2007bv}
C.~Burrage and A.~C. Davis, {\it {P}-term strings and semi-local strings},
  {\em JHEP} {\bf 11} (2007) 023,
  [\href{http://xxx.lanl.gov/abs/0707.3610}{{\tt 0707.3610}}].

\bibitem{Achucarro:2004ry}
A.~Achucarro and J.~Urrestilla, {\it {{F}-term strings in the Bogomolnyi limit
  are also {BPS} states}},  {\em JHEP} {\bf 08} (2004) 050,
  [\href{http://xxx.lanl.gov/abs/hep-th/0407193}{{\tt hep-th/0407193}}].

\bibitem{Davis:2005jf}
S.~C. Davis, P.~Binetruy, and A.-C. Davis, {\it {Local axion cosmic strings
  from superstrings}},  {\em Phys. Lett.} {\bf B611} (2005) 39--52,
  [\href{http://xxx.lanl.gov/abs/hep-th/0501200}{{\tt hep-th/0501200}}].

\bibitem{Achucarro:2006ef}
A.~Achucarro and K.~Sousa, {\it {A note on the stability of axionic {D}-term
  strings}},  {\em Phys. Rev.} {\bf D74} (2006) 081701,
  [\href{http://xxx.lanl.gov/abs/hep-th/0601151}{{\tt hep-th/0601151}}].

\bibitem{Witten:1981mf}
E.~Witten, {\it A simple proof of the positive energy theorem},  {\em Commun.
  Math. Phys.} {\bf 80} (1981) 381.

\bibitem{Edelstein:1995ba}
J.~D. Edelstein, C.~Nunez, and F.~A. Schaposnik, {\it Supergravity and a
  {B}ogomolnyi bound in three-dimensions},  {\em Nucl. Phys.} {\bf B458} (1996)
  165--188, [\href{http://xxx.lanl.gov/abs/hep-th/9506147}{{\tt
  hep-th/9506147}}].

\bibitem{Collinucci:2006sp}
A.~Collinucci, P.~Smyth, and A.~Van~Proeyen, {\it The energy and stability of
  {D}-term strings},  {\em JHEP} {\bf 02} (2007) 060,
  [\href{http://xxx.lanl.gov/abs/hep-th/0611111}{{\tt hep-th/0611111}}].

\bibitem{Wald:1984rg}
R.~M. Wald, {\em General Relativity}.
\newblock Chicago Univ. Pr., 1984.

\bibitem{Nester:1982tr}
J.~M. Nester, {\it A new gravitational energy expression with a simple
  positivity proof},  {\em Phys. Lett.} {\bf A83} (1981) 241.

\bibitem{Brax:2006yb}
P.~Brax, C.~van~de Bruck, A.~C. Davis, and S.~C. Davis, {\it Fermionic zero
  modes of supergravity cosmic strings},  {\em JHEP} {\bf 06} (2006) 030,
  [\href{http://xxx.lanl.gov/abs/hep-th/0604198}{{\tt hep-th/0604198}}].

\bibitem{Achucarro:1999it}
A.~Achucarro and T.~Vachaspati, {\it Semilocal and electroweak strings},  {\em
  Phys. Rept.} {\bf 327} (2000) 347--426,
  [\href{http://xxx.lanl.gov/abs/hep-ph/9904229}{{\tt hep-ph/9904229}}].

\bibitem{Taubes:1979tm}
C.~H. Taubes, {\it Arbitrary {N}-vortex solutions to the first order
  {G}inzburg-{L}andau equations},  {\em Commun. Math. Phys.} {\bf 72} (1980)
  277.

\bibitem{Hawking:1995fd}
S.~W. Hawking and G.~T. Horowitz, {\it The gravitational hamiltonian, action,
  entropy and surface terms},  {\em Class. Quant. Grav.} {\bf 13} (1996)
  1487--1498, [\href{http://xxx.lanl.gov/abs/gr-qc/9501014}{{\tt
  gr-qc/9501014}}].

\end{thebibliography}\endgroup

\end{document}